\documentclass[a4paper,10pt]{article}
\usepackage{epsfig}
\begin{document}

\title{Constraints from the old quasar APM 08279+5255 on two classes of $\Lambda(t)$-cosmologies}
\author{J. F. Jesus\thanks{E-mail: jfernando@astro.iag.usp.br}\\
Instituto de Astronomia, Geof\'{\i}sica e Ci\^{e}ncias Atmosf\'{e}ricas, USP,\\
Rua do Mat\~ao, 1226, S\~ao Paulo, SP, Brasil}

\maketitle

\begin{abstract} 
The viability of two different classes of $\Lambda(t)$CDM 
cosmologies is tested by using the APM 08279+5255, an old quasar 
at redshift $z = 3.91$. In the first class of models, the 
cosmological term scales as $\Lambda(t)\sim R^{-n}$. The 
particular case $n=0$ describes the standard $\Lambda$CDM model 
whereas $n=2$ stands for the Chen and Wu model. For an estimated 
age of 2 Gyr, it is found that the power index has a lower limit 
$n> 0.21$, whereas for 3 Gyr the limit is  $n 
> 0.6$. Since $n$ can not be so large as $\sim 0.81$, the 
$\Lambda$CDM and Chen and Wu models are also ruled out by this 
analysis. The second class of models is the one recently proposed 
by Wang and Meng which describes several $\Lambda(t)$CDM 
cosmologies discussed in the literature. By assuming that the true 
age is  2 Gyr it is found that the $\epsilon$ parameter satisfies 
the lower bound $\epsilon 
> 0.11$, while for 3 Gyr, a lower limit of $\epsilon 
> 0.52$ is obtained. Such limits  are 
slightly modified when the baryonic component is included. 
\end{abstract} 

\section{Introduction} 

Many cosmological models driven by dark energy have been proposed 
in the literature for explaining the discovered cosmic 
acceleration of the Universe \cite{perlmutter}. At present, the 
preferred scenario is provided by the so-called $\Lambda$CDM 
model \cite{Lima}. However, this model is plagued with the 
Cosmological Constant Problem (CCP) \cite{weinberg}. Such a problem 
occurs because the value of the Cosmological Constant (CC) 
suggested by the recent observations is incredibly small as 
compared to simple estimates from Quantum Theory of Fields (about 
120 orders of magnitude smaller). 

Decaying Vacuum Cosmologies or $\Lambda$(t)-Cosmologies \cite{Ozer} 
is a possibility to alleviate this problem. This kind of models 
try to explain the CCP through the hypothesis that $\Lambda$ 
couples with matter in such a way that it decays with time. 
Qualitatively, this means that the value of $\Lambda$ is very 
small today because the Universe is too old. 

$\Lambda$(t)-cosmologies are defined by some direct (or indirect) 
phenomenological time dependence of $\Lambda$ whose free 
parameters must be constrained by the cosmological tests. One of 
these tests is the age of the universe at high 
redshifts \cite{alc99}.  Since the Universe must be older than any 
object contained in it, the basic idea of this test is to compare 
the age of the Universe at a given redshift with the estimated age 
of the oldest objects at the same redshift. 

In this paper, the age test is applied for two classes of 
$\Lambda$(t)-cosmologies. The first one generalizes the Chen and 
Wu model \cite{cw} by considering a cosmological term scaling as 
$\Lambda(t) \sim R^{-n}$. The second class is the one recently 
proposed by Wang and Meng \cite{wm}. In this scenario, the energy 
density of the CDM component scales as $\rho_{m} \sim 
R^{-3+\epsilon}$.  The prediction of both models are confronted to 
the old quasar APM 08279+5255 at redshift z = 3.91, whose age was 
recently estimated trough its iron 
abundance \cite{hasinger,friaca}. 

\section{Extended Chen and Wu Model} 
In 1990, by using dimensional arguments, Chen and Wu \cite{cw} 
proposed the following functional form for the cosmological term: 

\begin{equation} 
\label{Eq:1} \Lambda = \frac{\alpha}{R^{2}}, 
\end{equation} 
where $\alpha$ is a dimensionless constant and $R(t)$ is the scale 
factor of the FRW type geometries. The power $n=2$ was fixed by 
considering that gravity is described by a classical field  after 
the Planck age. In what follows, by assuming that the vacuum 
energy density, and the cosmic acceleration at the present stage 
may have a quantum origin, we discuss the general $\Lambda(t)$ 
dependence \cite{jackson} 

\begin{equation} 
\label{Eq:2} \Lambda = \frac{\alpha}{R^{n}}. 
\end{equation} 
An advantage of this functional form is that the Chen and Wu model 
is recovered  for $n = 2$ whereas, for $n=0$, it reduces to the 
cosmic concordance ($\Lambda$CDM) model. 

Now, by assuming spatial flatness as predicted by inflation and 
observationally suggested by the WMAP experiments \cite{wmap}, the 
FRW equation plus the energy conservation law can be written as 

\begin{eqnarray} 
8\pi G\rho_m +  \frac{\alpha}{R^{n}}= 3H^{2},\label{Eq:3}\\ 
\frac{d}{dR}\left(\rho_m R^3\right)=\frac{n\alpha}{8\pi 
GR^{n-2}},\label{Eq:4} 
\end{eqnarray} 
where $\rho_m$ is the energy density for the pressureless cold 
dark matter (CDM) fluid, and $H={\dot R}/R$ is the Hubble 
parameter. 

Now, let us determine the expression for $t_z$, the age of the 
Universe at redshift $z$. First, one must integrate the energy 
conservation law for obtaining the $\rho_m$, and inserting the 
result into the FRW equation, it thus follow that 

\begin{equation} 
\label{Eq:12} 
H^2=H_0^2\left[\left(\frac{n-3\Omega_{mo}}{n-3}\right)\left(\frac{R_o}{R}\right)^3+ 
\frac{3\Omega_{\Lambda0}}{3-n}\left(\frac{R_o}{R}\right)^n\right]. 
\end{equation} 

Note that for $n=0$ this expression for the Hubble parameter 
reduces to the one of the flat $\Lambda$CDM model, as should be 
expected. Finally, by integrating again, we have for $t_z$: 

\begin{equation} 
\label{Eq:13} t_z=H_0^{-1}\int_{0}^{(1+z)^{-1}} \left[ 
\left(\frac{n-3\Omega_{mo}}{n-3}\right)x^{-1}+\frac{3(1-\Omega_{mo})x^{2-n}}{(3-n)}\right]^{-\frac{1}{2}}dx, 
\end{equation} 
where $x=R/R_o$ is a convenient integration variable. It is 
straightforward to find similar expressions for the logarithm 
solution ($n = 3$), but this case will not be considered here. It 
can also be shown that $n$ can not be so large as one wishes. 
Actually, the weak energy condition for the matter density implies 
an upper bound  $n \sim 0.81$ (see Appendix). 

\section{The Wang and Meng (WM) Model} 
More recently, a new  decaying vacuum cosmology was proposed by 
Wang and Meng \cite{wm}. Unlike the common approach in the 
literature, they did not assume a functional form for 
$\Lambda$(t). In such a scenario, the decay law is deduced from 
its effect on the CDM evolution. Qualitatively, since the vacuum 
is decaying into CDM particles, the created matter must dilute 
more slowly in comparison with the standard evolution ($\rho_m 
\propto R^{-3}$). Thus, assuming that the deviation from the 
conservation law is characterized by a positive constant 
$\epsilon$, one may write 
\begin{equation} 
\label{Eq:15} \rho_{m} = 
\rho_{mo}\left(\frac{R}{R_o}\right)^{-3+\epsilon}, 
\end{equation} 
where $\epsilon$ is a small positive constant and $\rho_{mo}$ is 
the present value of $\rho_{m}$. Inserting the above expression in 
the energy conservation law, one finds 
\begin{equation} 
\label{Eq:16} 
\left(\frac{R_o}{R}\right)^{3}\frac{d}{dR}\left(\rho_{mo} 
\left(\frac{R}{R_o}\right)^\epsilon\right)=-\frac{d\rho_{\Lambda}}{dR}. 
\end{equation} 
Hence, by integrating the above expression, it thus follows that 

\begin{equation} 
\rho_{\Lambda} = {\tilde{\rho}}_{\Lambda0} + 
\frac{\epsilon\rho_{mo}}{3-\epsilon}\left(\frac{R}{R_o}\right)^{-3+\epsilon}, 
\end{equation} 
where ${\tilde{\rho}}_{\Lambda0}$ is an integration constant which 
can be  associated to ``the ground state of vacuum" (note that the 
present vacuum energy density is  $\rho_{\Lambda0} = 
{\tilde{\rho}}_{\Lambda0} + {\epsilon\rho_{mo}}/{(3-\epsilon)}$). 

Under such conditions, the FRW equation for the WM model reads: 

\begin{equation} 
\label{Eq:19} H^{2} = \frac{8\pi G}{3}(\rho_{m} + \rho_{\Lambda}) 
= 
H^{2}_0\left[\frac{3\Omega_{mo}}{3-\epsilon}\left(\frac{R_o}{R}\right)^{3-\epsilon} 
+ \tilde{\Omega}_{\Lambda0}\right], 
\end{equation} 
with an  age-redshift relation given by 

\begin{equation} 
\label{Eq:22} t_{z} = H^{-1}_0\int_{0}^{(1+z)^{-1}} 
\frac{dx}{\sqrt{(1 - {\tilde{\Omega}}_{\Lambda0})x^{\epsilon - 1} 
+ {\tilde{\Omega}}_{\Lambda0}x^{2}}}\,. 
\end{equation} 
As should be expected, by taking $\epsilon = 0$ in the above 
expression (no CDM creation), the age-redshift relation for the 
$\Lambda$CDM model is recovered. 

\section{The Age-Redshift Test} 
Let us now discuss some constraints by considering the quasar APM
08279+5255, as a cosmic clock. Such a quasar, located at $z=3.91$,
has an estimated age from 2 to 3 Gyr, with a best fit age 
of 2.1 Gyr \cite{hasinger,friaca}. As first discussed by Alcaniz 
and Lima \cite{alc99}, the age test here is defined by the 
condition 

\begin{equation} 
\label{Eq:23} t_z \geq t_g, 
\end{equation} 
where $t_z$ is the age of the Universe at redshift $z$ and $t_g$ 
is the age of the quasar at the same redshift. It just says that 
the Universe is older than any structure that it contains. 

For test both models we have used the optimal value of the matter 
density parameter as given by the WMAP team \cite{wmap}, 
$\Omega_{mo} = 0.27 \pm 0.04$, whereas the Hubble parameter is the 
one provided by the HST team \cite{hst}, $H_0 = 72 \pm 8 
kms^{-1}Mpc^{-1}$. It is convenient to introduce a dimensionless 
age parameter, $T_g=H_0t_g$, in such a way that for the 2.0 Gyr 
old quasar, this value of the Hubble parameter constrains it to 
the range $0.131 \leq T_g \leq 0.163$. It follows that $T_g \geq 
0.131$. Therefore, for a given value of $H_0$, only models with an 
expanding age bigger than this value at $z = 3.91$ will be 
compatible with the existence of this object. In order to assure 
the robustness of our analysis, it is also necessary to adopt the 
lower bound for the value of the Hubble parameter, $H_0 = 
64kms^{-1}Mpc^{-1}$. 
\subsection{Constraints on Chen and Wu Model} 
The results for the Chen and Wu model can be seen on 
Table~\ref{tab1}. The constraints on the power $n$ are heavily 
dependent on the estimated age of the quasar. By considering that 
the correct age is 2 Gyr, the power $n$ has a lower limit given by 
$n\geq 0.21$. This minimal value increases for $0.28$ when we 
consider the optimal estimated age of 2.1 Gyr and it is $n>0.6$ if 
the age is $3.0$ Gyr. In particular, these results imply that the 
$\Lambda$CDM ($n = 0$) is ruled out by this analysis, if we have 
$\Omega_{\Lambda0} = 0.73$. This result agrees with the ones 
recently determined by some authors for the standard $\Lambda$CDM 
model using the same quasar. They found that $\Omega_{\Lambda0} 
\geq 0.78$ in order to be compatible with the existence of this 
quasar \cite{alc03}.

\begin{table}[htbp]{
\begin{center}
\begin{tabular}{cc}
\hline
\multicolumn{2}{c}{Table 1. Values of $n_{min}$ for the}\\
\multicolumn{2}{c}{three estimated ages of quasar}\\
\hline
\hspace{0.5cm}$t_g$ & \hspace{1cm}$n_{min}$\\
\hspace{0.5cm}(Gyr) \\ 
\hline
\hspace{0.5cm}2 & \hspace{1cm}0.21\\ 
\hspace{0.5cm}2.1 & \hspace{1cm}0.28\\ 
\hspace{0.5cm}3 & \hspace{1cm}0.60\\
\end{tabular}\label{tab1}
\end{center}} 
\end{table} 

\begin{figure}[!htb] 
\centerline{\epsfig{file=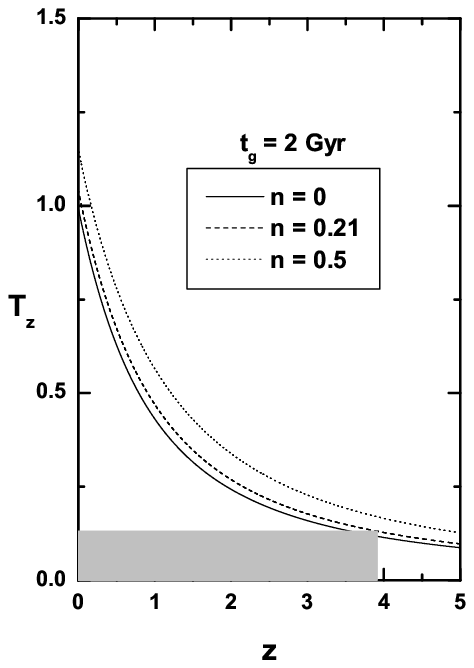,width=5.5cm,height=6.5cm}\epsfig{file=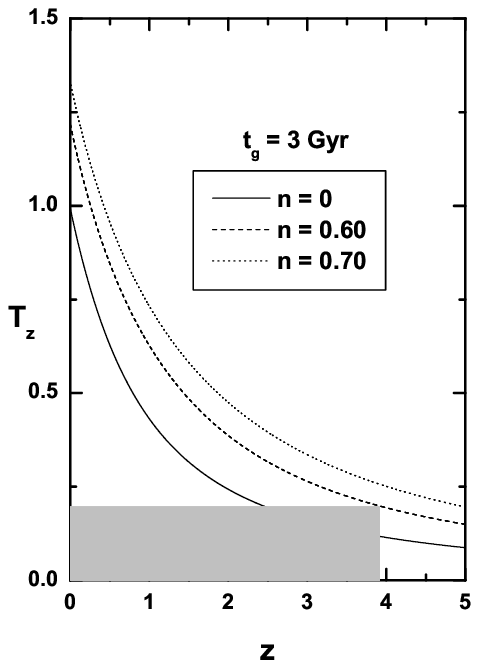,width=5.5cm,height=6.5cm}} 
\caption{$t_z${\it versus} redshift for two estimated ages of the quasar. Left 
panel: 2 Gyr. Right panel: 3 Gyr. The curves correspond to each 
value of {\it n} as indicated. The rectangle corresponds to the 
age of the quasar.\label{fig_nmin}} 
\end{figure} 

\subsection{Constraints on Wang and Meng Model} As in the previous model, we have made 
use of a routine in C to perform the age test for the WM model. 
The main results of our analysis are displayed in 
Table~\ref{tab2}. As can be seen there, the minimal value of the 
free parameter $\epsilon$ is modified when the baryonic component 
is included. This kind of model was recently considered by Alcaniz 
and Lima \cite{alcl05}. Note that the minimum value of $\epsilon$ 
varies in the range $0.115 < \epsilon < 0.527$. 

\begin{table}[htbp] 
{\begin{center}
\begin{tabular}{ccc}
\hline
\multicolumn{3}{c}{Table 2. Values of $\epsilon_{min}$ for the three estimated}\\
\multicolumn{3}{c}{ages of quasar (with and without baryons)}\\
\hline
$t_g$ & \hspace{0.5cm}$\epsilon_{min}$ & $\epsilon_{min}$ \\
(Gyr) & \hspace{0.5cm}(no baryons) & (with baryons) \\
\hline
 2 & \hspace{0.5cm}0.115 & 0.231 \\
 2.1 & \hspace{0.5cm}0.163 & 0.296 \\
 3 & \hspace{0.5cm}0.527 & 0.910 \\
\end{tabular} \label{tab2}
\end{center}} 
\end{table} 

It is also interesting to know how the minimal value of $\epsilon$ 
depends on matter density parameter. In  Figure 2, the values of 
$\epsilon_{min}$ are plotted as a function of $\Omega_{mo}$ for 
the WM model with no baryons. We see that $\epsilon_{min}$ is 
heavily dependent on the values of $\Omega_{mo}$. It becomes 
negative for $\Omega_{mo} < 0.2$ whether the age of the quasar is 
2 Gyr. The lower limit is nearly $0.1$ for 3 Gyr. It is worth 
notice that negative values of $\epsilon$ are thermodynamically 
forbidden \cite{alcl05}.

\begin{figure}[htpb] 
\centerline{\epsfig{file=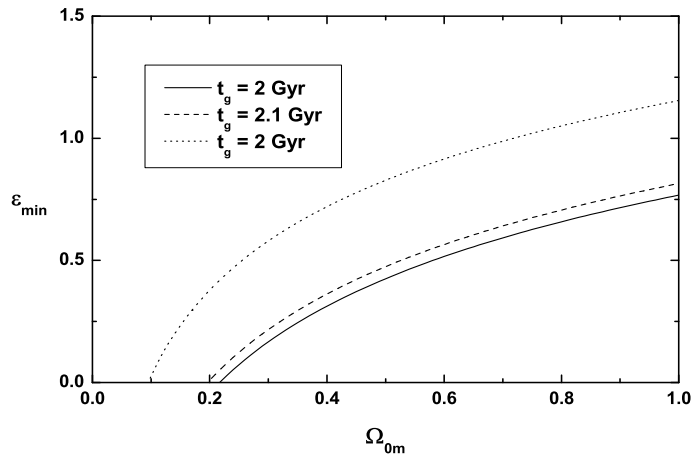,width=10cm,height=7cm}} \caption{The 
parameter  $\epsilon_{min}$ {\it versus} $\Omega_{mo}$ for the WM 
model with no baryons. As indicated in the panel,  3 different 
estimated ages of the quasar has been considered.\label{fig3}} 
\end{figure} 

Alcaniz and Lima \cite{alcl05} also analyzed the WM model using Sne 
Ia, X-Ray luminosity from galaxy clusters and CMB data. A best fit 
of $\epsilon = 0.06 \pm 0.10$ at 95\% c.l. was established in 
their joint analysis. Concerning the present age test, this result 
is not compatible with the existence of the quasar APM 08279+5255 
(see Table 2). Actually, even for an estimated age of 2.1 Gyr, the 
best fit found on Ref.~ \cite{friaca}, the above value disagrees 
from our minimal value  by 3 standard deviations. Naturally, if 
the real age is 3 Gyr the situation becomes worst. Although 
improving the determination of the critical redshift, the presence 
of a baryonic component worsens the scenario for the age test 
because it decelerates the Universe thereby decreasing the age at 
a given redshift. 
\section{Conclusions} 
The age test has been applied for two different $\Lambda(t)$ 
cosmologies by using the old quasar APM 08279+5255. For $\Lambda 
\sim R^{-n}$ (extended Chen and Wu model) we found $n\geq 0.21$. 
This means that the cosmic concordance $\Lambda$CDM model ($n=0$) 
is incompatible with the existence of this object. This result is 
in line with the previous analysis by Alcaniz et al. \cite{alc03}. 
However, there is an upper limit on $n$ from physical and 
observational considerations. In particular, the original CW model 
($n=2$) is not compatible with the WMAP data because it requires 
$\Omega_{mo} > 2/3$ (see also Appendix). 

Finally, we have established new limits to the $\epsilon$ 
parameter of the Wang and Meng model (with and with no baryons) 
which are more stringent than the ones recently determined by Alcaniz and Lima \cite{alcl05}. 
Until the present, the existence of the 
old quasar  constrains severely all the models present in the 
literature. 

\section*{Acknowledgments} 

This work was supported by CNPq (Brazilian Research Agency). I 
would like to thank Prof. J. Ademir S. Lima for suggesting the 
problem. I am also grateful to Rodrigo Holanda and Rose Santos for 
the encouragement and helpful discussions. 

\appendix 
\section{Extended CW model and its maximum power $n$} 

In the extended CW model, the matter density reads (see Eqs. 
(3)-(5)): 

\begin{equation} 
\label{Eq:A1} \rho_m = \frac{3H_0^2}{8\pi 
G}\left[\frac{n-3\Omega_{mo}}{n-3}\left(\frac{R_o}{R}\right)^{3}+ 
\frac{n\Omega_{\Lambda0}}{(3-n)}\left(\frac{R_o}{R}\right)^n\right]. 
\end{equation} 
Now, supposing $n < 3$, it is easy to see that the matter density 
is always positive only if  $n \leq 3\Omega_{mo}$. Thus, $n$ 
should be less than $\sim 0.81$ for $\Omega_{mo} = 0.27$ as 
suggested from CMB experiments \cite{wmap}. Actually, at $2\sigma$, 
the maximum value allowed for $n$ is $\sim 1.05$ for $\Omega_{mo} 
= 0.35$. Hence, we may conclude that the original CW model ($n = 
2$) is ruled out by the WMAP data.

\end{document}